# Theoretical Puncture Mechanics of Soft Compressible Solids

Stefano Fregonese, Zhiyuan Tong, Sibo Wang, Mattia Bacca[*]

*Mechanical Engineering Department, institute of Applied Mathematics, School of Biomedical Engineering, University of British Columbia, Vancouver BC V6T1Z4, Canada*

[*]Corresponding author. *E-mail address:* mbacca@mech.ubc.ca

**Abstract**

Accurate prediction of the force required to puncture a soft material is critical in many fields like medical technology, food processing, and manufacturing. However, such a prediction strongly depends on our understanding of the complex nonlinear behavior of the material subject to deep indentation and complex failure mechanisms. Only recently we developed theories capable of correlating puncture force with material properties and needle geometry. However, such models are based on simplifications that seldom limit their applicability to real cases. One common assumption is the incompressibility of the cut material, albeit no material is truly incompressible. In this paper we propose a simple model that accounts for linearly elastic compressibility, and its interplay with toughness, stiffness, and elastic strain-stiffening. Confirming previous theories and experiments, materials having high-toughness and low-modulus exhibit the highest puncture resistance at a given needle radius. Surprisingly, in these conditions, we observe that incompressible materials exhibit the lowest puncture resistance, where volumetric compressibility can create an additional (strain) energy barrier to puncture. Our model provides a valuable tool to assess the puncture resistance of soft compressible materials and suggests new design strategies for sharp needles and puncture-resistant materials.

*Keywords*: *Puncture; Piercing; Compressibility; Bulk Modulus; Cutting; Soft Materials*

**Introduction**

Puncture mechanics in soft solids has recently gained significant attention due to its extensive range of engineering and biomedical applications. The success and safety of delicate operations such as tissue biopsy [1-2], blood sampling [3], drug injection [4-5], and surgical intervention [6], require a precise prediction of the puncture resistance of soft tissue. However, such a prediction is often empirical and, thus, subject to human error. This limitation becomes important especially when the abovementioned medical operations are automated [7]. This highlights the need for a theory that allows accurate and quantitative predictions of the force required to pierce a material, from its mechanical properties. Puncture and cutting mechanics are also parts of several natural processes in the living kingdom, where tissue piercing is a necessity [8-11], as well as engineering applications involving food processing [12-13], manufacturing [14-15], and (in-situ) material testing [16-17]. The latter is of growing importance due to the complication emerging in the attempt to characterize the fracture resistance of soft polymeric gels [18-20].

The first theoretical model describing the mechanics of soft solid puncture was provided by [21], and later validated by the same authors [22] by puncturing rubber and porcine skin. This model correlates the fracture toughness and shear modulus of the cut material, as well as the needle radius, with the force required to penetrate a material, once the needle is inserted. The model described the penetration mechanics of blunt flat-head cylindrical indenters (needles), as well as for sharp conical-tip cylindrical indenters. In the former case, the indenter would penetrate the material by propagating a cylindrical crack in Mode II, while in the latter, the penetration would occur via the propagation of a planar crack in Mode I. These authors, however, did not provide a prediction for the needle insertion force, commonly generating a peak in the force-displacement curve. Moreover, the proposed model assumes frictionless needle-specimen contact. [23-24] provided an experimental investigation of the correlation between needle radius and material properties in defining the needle insertion force. [25] extended the model by [21] to predict the critical force and depth at needle insertion, validated against the experiments by [23-24]. The models proposed by [21] and [25], as well as the asymptotic linear elastic simplification proposed by [26], relied on the hypothesis of frictionless contact between needle and specimen. The experimental validation of [25]'s model suggests that friction does not play a major role at needle insertion, while it becomes very important after insertion, during needle penetration at significant penetration depth. This was proven by [27], who extended the theory by [21] to incorporate the effect of friction and adhesion, and validated their new model against experiments from [22] and [24]. The frictionless prediction of the penetration force, however, can be assumed as valid at the onset of penetration, suddenly after the needle is inserted into the specimen, and at shallow penetration depths.

All the above models rely on the hypothesis of incompressible or nearly-incompressible materials, *i.e.*, where the volumetric strain of the cut material is neglected. Most materials, however, exhibit volumetric compressibility. *I.e.*, they experience volumetric strain. In this paper we propose a model that accounts for the volumetric compressibility of soft materials and its role in defining puncture resistance. Our model assumes quasi-static puncture, *i.e.* the needle moves at very low velocity, and linear elastic volumetric compressibility. Also, the needle is modeled as a rigid cylindrical indenter having a spherical tip. Confirming previous observations, we observe that tough and/or soft materials exhibit the highest puncture resistance (highest puncture force for a given needle radius). Surprisingly, volumetric compressibility renders these materials even more puncture resistant. We further explore puncture mechanics looking at the interplay between strain stiffening, toughness, stiffness, and volumetric compressibility. The latter is described as the ratio of bulk-to-shear moduli $\kappa/\mu$ (incompressible materials having $\kappa \gg \mu$).

**Model, Results and Discussion**

The model system is sketched in Figure 1. Here, a cylindrical needle of radius $R$, having a spherical tip, is pushed against a specimen of soft material by a depth $d$, requiring a force $F$. As described by [25], when the depth is shallower than a critical $d_c$, *i.e.* $d < d_c$, the needle indents the material without rupturing it. At the critical depth, $d = d_c$, the force is now the critical $F_c$, and the needle enters the specimen by fracturing it. Once the needle is inserted into the specimen, $d > d_c$, the needle penetrates the specimen, since this process is energetically favoured over indentation. In this study we assume frictionless conditions between needle and specimen, as adopted by [Shegold2004, Shegold2005, 25]. Under this hypothesis, the needle penetrates the specimen at the constant force $F_p$. As observed by [27], most experiments involve friction, and the latter induced a linear increment of penetration force. Thus, the hypothesis of frictionless conditions can provide a good estimation only at shallow penetration depths after insertion.

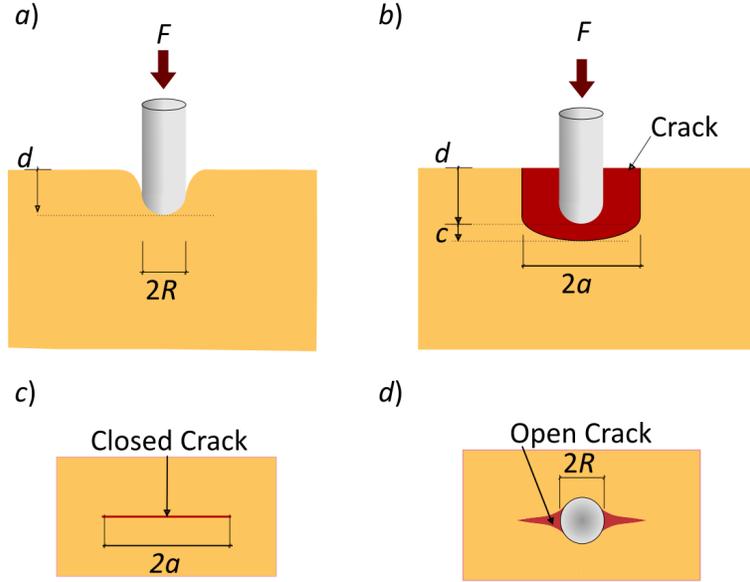

**Figure 1**: Schematics of the model system. The needle of radius $R$ indents the material by depth $d$ *a)* and later penetrates the material by nucleating, propagating, and opening a crack of undeformed half width $2a$ to accommodate the needle *b-d)*. In the latter case, we report the side view *b)* and top view *d)*, as well as the top view of the closed crack after the needle is extracted from the specimen *c)*.

The hyperplastic behaviour of the soft material is modelled as a *linearly compressible* 1-term Ogden [28] material, which strain energy density $\psi$ is given by

$$\frac{\psi}{\mu} = \frac{\psi_D}{\mu} + \frac{\psi_V}{\mu} \qquad (1a)$$

with

$$\frac{\psi_D}{\mu} = \frac{2}{\alpha^2}\left(\bar{\lambda}_1^\alpha + \bar{\lambda}_2^\alpha + \bar{\lambda}_3^\alpha - 3\right) \qquad (1b)$$

the (dimensionless) *deviatoric* contribution, and

$$\frac{\psi_V}{\mu} = \frac{\kappa}{\mu}\frac{1}{2}(J-1)^2 \qquad (1c)$$

the (dimensionless) *volumetric* one. In Eq. (1), $\bar{\lambda}_i = \lambda_i J^{-1/3}$ is the deviatoric principal stretch in the direction $i = 1,2,3$, with $\lambda_i$ the corresponding principal stretch, and $J = dV/dV_0$ is the swelling ratio, where $dV$ and $dV_0$ are the unit volumes in the current and reference states, respectively. In the same equations, $\mu$ and $\kappa$ are the shear and bulk moduli of the material (making $\psi/\mu$ dimensionless), and $\alpha$ is the strain-stiffening coefficient of the material, where $\alpha = 2$ recovers neo-Hookean elasticity. Eq. (1) shows that the relation between dimensionless strain energy density $\psi/\mu$ and the deformation state, defined by $\bar{\lambda}_i$ and $J$, depends on (*I*) the dimensionless *strain-stiffening coefficient* $\alpha$ and (*II*) the *dimensionless bulk modulus* $\kappa/\mu$. As we will see further, these two parameters control the configuration of the needle-specimen system and thus affect the critical indentation depth $d_c$ and force $F_c$ at needle insertion, and the required force for needle penetration $F_p$.

The principal Cauchy (true) stress in direction 1, $\sigma_1$, is given by

$$\frac{\sigma_1}{\mu} = \frac{\lambda_1}{J}\frac{\partial}{\partial \lambda_1}\left(\frac{\psi}{\mu}\right) \tag{2a}$$

with $\psi/\mu$ taken from Eq. (1). Eq. (2a) rewrites to

$$\frac{\sigma_1}{\mu} = \frac{\sigma_{1D}}{\mu} - \frac{p}{\mu} \tag{2b}$$

where $J\sigma_{1D} = \lambda_1 \partial \psi_D/\partial \lambda_1$, giving

$$\frac{\sigma_{1D}}{\mu} = \frac{2}{3\alpha J}\left(2\bar{\lambda}_1^\alpha - \bar{\lambda}_2^\alpha - \bar{\lambda}_3^\alpha\right) \tag{2c}$$

the *deviatoric* stress $\sigma_{1D}$, and $p = -\partial \psi_V/\partial J$, giving

$$\frac{p}{\mu} = -\frac{\kappa}{\mu}(J-1) \tag{2d}$$

the *hydrostatic* pressure $p$ sustained by the material. The same stress in directions 2 and 3 can be obtained by substituting $\bar{\lambda}_1$ with $\bar{\lambda}_2$ and $\bar{\lambda}_3$, respectively, in Eq. (2). The hydrostatic pressure $p = (\sigma_1 + \sigma_2 + \sigma_3)/3$ is linearly proportional to the volumetric swelling, as shown in Eq. (2d), highlighting the adopted linear elastic description of volumetric compressibility. Following the theory proposed by [25], we will now define the two needle-specimen configurations, namely (*i*) indentation and (*p*) penetration.

*Needle Indentation*

Prior to needle insertion, *i.e.* during needle indentation, the force $F_i$ applied to the back of the needle to achieve an indentation depth $d$ is calculated via finite element analysis (FEA) and fitted to the law [25],

$$\frac{F_i}{\mu R^2} = B\left(\frac{d}{R}\right)^b \tag{3}$$

to the least square fits with an accuracy of $r^2 = 0.9997 - 0.9999$ within a maximum indentation depth of $d = 4R$. In Eq. (3), $B$ and $b$ are dimensionless parameters that depend on the dimensionless *strain-stiffening coefficient* $\alpha$ and on the *dimensionless bulk modulus* $\kappa/\mu$, as given in Table 1. In this table, the dimensionless bulk modulus is also expressed as a function of the Poisson's ratio of the undeformed material, where $\kappa/\mu = 2(1+\nu)/3(1-2\nu)$ is inverted to give

$$\nu = \frac{3\kappa - 2\mu}{3\kappa + \mu} \tag{4}$$

The work done by the needle is $w_i = \int_0^d F_i \delta d$, thus giving

$$\frac{w_i}{\mu R^3} = \frac{B}{b+1} \left(\frac{d}{R}\right)^{b+1} \tag{5}$$

**Table 1**: Parameters value for Eq. (3) and (5), namely $B$ and $b$ as functions of the strain-stiffening coefficient $\alpha$, and the dimensionless bulk modulus $\kappa/\mu$ (or Poisson's ratio $\nu$, Eq. (4)) of the material

| $\alpha$ | $\kappa/\mu$ ($\nu$) | $B$ | $b$ |
|---|---|---|---|
| 3 | 0.0667 (−0.75) | 2.411 | 1.553 |
|   | 0.1667 (−0.5)  | 2.652 | 1.523 |
|   | 1.667 (0.25)   | 3.533 | 1.474 |
| 5 | 0.0667 (−0.75) | 2.813 | 1.488 |
|   | 0.1667 (−0.5)  | 2.875 | 1.522 |
|   | 1.667 (0.25)   | 3.023 | 1.582 |

*Needle Penetration*

Once the needle is inserted into the specimen, it will penetrate the material. The required mechanical work, $w_p$, must compensate for the energetic requirement for crack nucleation and propagation, $2G_c a d$, and for crack opening $S_p d$, giving

$$\frac{w_p}{\mu R^3} = 2 \frac{G_c}{\mu R} \frac{a}{R} \frac{d}{R} + \frac{S_p}{\mu R^2} \frac{d}{R} \tag{6}$$

Here, $G_c$ is the work of fracture (toughness) of the material, $a$ is the undeformed half-width of the crack (see Figure 1), and $S_p$ is the mechanical work, per unit penetration depth, required to open the crack to accommodate the presence of the needle (spacing energy).

The penetration force $F_p = \partial w_p / \partial d$ can be calculated from Eq. (6) as

$$\frac{F_p}{\mu R^2} = 2 \frac{G_c}{\mu R} \frac{a}{R} + \frac{S_p}{\mu R^2} \tag{7}$$

In Eq. (7), $G_c/\mu R$ is the *relative toughness* of the material, while $\mu R/G_c$ is the *relative radius* of the needle. This parameter controls the critical needle-specimen configuration, given by the ratio $a/R$ (see Figure 1), together with the *strain-stiffening coefficient* $\alpha$, and the *dimensionless bulk modulus* $\kappa/\mu$ of the material. Figure 2 shows the model adopted for our FEA, where we select a crack size $a$ and needle radius $R$, thus defining the ratio $a/R$, and compute the J-integral of the material around the crack tip and the strain energy $S_p$ accumulated in the material by pushing the needle into the quarter specimen (Figure 2). The calculated J-integral corresponds to the energy release rate $G$ at the crack tip. Thus, by imposing $G = G_c$, the selected ratio $a/R$ corresponds to the critical needle-specimen configuration. The correlation between the dimensionless J-integral, *i.e.* the relative toughness $G_c/\mu R$ of the material, and the ratio $a/R$ is reported in Figure 3a, as a function of the *strain-stiffening coefficient* $\alpha$, and the *dimensionless bulk modulus* $\kappa/\mu$ of the material. This relation is then fitted to the function

$$\frac{a}{R} = \frac{a_0}{R} + Q\left(\frac{\mu R}{G_c}\right)^q \tag{8}$$

to the least square fits with an accuracy of $r^2 = 0.996 - 0.999$, where $a_0/R$ is the limit for $\mu R/G_c = 0$, and $Q$ and $q$ are dimensionless constants reported in Table 2, as functions of $\alpha$ and $\kappa/\mu$. It should be noted that we expect $a_0/R \approx 0$, since $\mu R/G_c = 0$ would yield the condition of cavitation, and this hypothesis is confirmed in Table 2, for $\alpha = 5$. The non-zero $a_0/R$ emerges from the mathematical convenience of fitting the plots in Figure 3a with a simple law like that in Eq. (8), which makes use of the minimal number of parameters.

From each FEA, with selected crack size $a/R$, the calculated dimensionless strain (spacing) energy $S_p/\mu R^2$ is plotted in Figure 3b against the calculated dimensionless J-integral, i.e. $G_c/\mu R$. The correlation between $S_p/\mu R^2$ and $a/R$ is fitted against the simple law

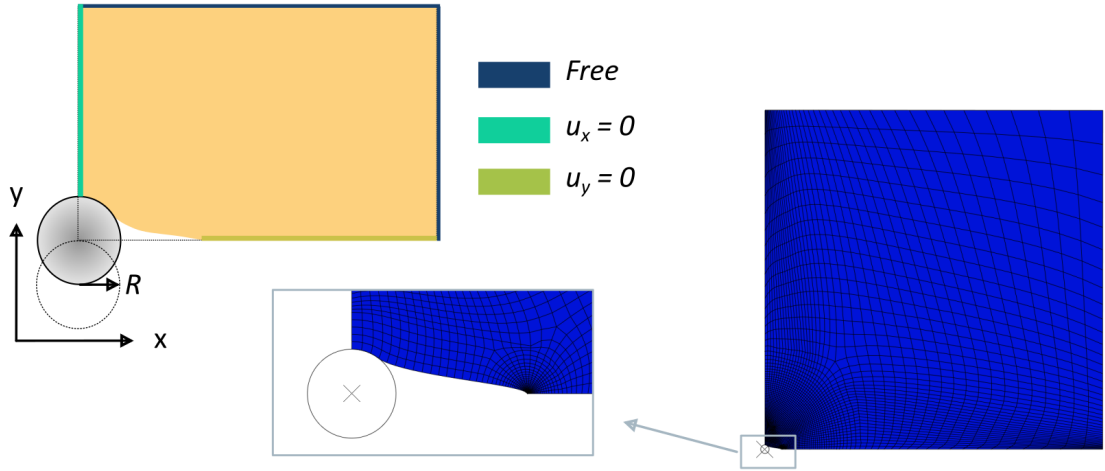

**Figure 2**: Schematics of the finite element model (left) and the adopted mesh (right)

$$\frac{S_p}{\mu R^2} = \frac{S_{p_0}}{\mu R^2} + M\left(\frac{R}{a}\right)^m \tag{9}$$

to the least square fits with an accuracy of $r^2 = 0.998 - 0.999$, where $S_{p_0}/\mu R^2$ is the limit for $R/a \to 0$, and $M$, and $m$ are dimensionless constants reported in Table 2, as functions of $\alpha$ and $\kappa/\mu$, for compressible materials, while for compressible materials these parameters are reported in [25]. It should be noted that the critical ratio $a/R$ shown in Figure 3a, and predicted by Eq. (8), is that at which further crack propagation (increasing $a/R$) would reduce the energy release rate below the critical $G_c$, hence crack propagation would stop, while crack closure (reduction of $a/R$) would increase the energy release rate above $G_c$, thus prompting crack propagation. Hence, the equilibrium condition at the critical $a/R$ is stable.

**Table 2**: Parameters value for Eq. (8) and (9), namely $B$ and $b$ as functions of the strain-stiffening coefficient $\alpha$, and the dimensionless bulk modulus $\kappa/\mu$ (or Poisson's ratio $\nu$) of the material. For incompressible materials, $\kappa/\mu \to \infty$, these parameters are reported in [25]

| $\alpha$ | $\kappa/\mu$ ($\nu$) | $a_0/R$ | $S_{p_0}/\mu R^2$ | $Q$ | $M$ | $q$ | $m$ | $q \cdot m$ |
|---|---|---|---|---|---|---|---|---|
| 3 | 0.0667 (−0.75) | 0.8261 | 0.5141 | 0.3048 | 1.4190 | 0.6510 | 2.4980 | 1.6262 |
|   | 0.1667 (−0.5)  | 0.7904 | 0.5644 | 0.3503 | 1.4450 | 0.6310 | 2.3000 | 1.4513 |
|   | 1.667 (0.25)   | 0.5817 | 0.9141 | 0.5079 | 1.5730 | 0.6210 | 1.4960 | 0.9290 |
| 5 | 0.0667 (−0.75) | 0 | 0.5012 | 1.0510 | 1.6890 | 0.3753 | 2.5730 | 0.9656 |
|   | 0.1667 (−0.5)  | 0 | 0.5632 | 1.0740 | 1.7230 | 0.3969 | 2.4180 | 0.9597 |
|   | 1.667 (0.25)   | 0 | 1.0260 | 1.1170 | 1.8500 | 0.4192 | 2.1930 | 0.9193 |

Figure 3a shows that the dimensionless crack size $a/R$ is inversely proportional to the relative toughness $G_c/\mu R$, confirming the findings of [21]. Because $G_c/\mu R$ controls the energetic ratio of crack-propagation to crack-opening, a higher $G_c/\mu R$ prompts the formation of a small crack that is then opened wide to accommodate the needle.

At high $G_c/\mu R$ we can observe that $a/R$ is almost independent of the dimensionless bulk modulus $\kappa/\mu$, and mainly controlled by strain-stiffening $\alpha$, particularly for $\alpha = 5$. From this, and also considering that in this regime $a/R$ is smallest, we can deduce that deviatoric deformation prevails over the volumetric one, in the material. We can also observe that $a/R$ is proportional to $\alpha$, since strain-stiffening increments the crack opening energy, thereby motivating an enlargement of the crack width $a/R$. Interestingly, we can observe that the bulk modulus $\kappa/\mu$ is inversely proportional to $a/R$, where incompressible materials achieve the smallest crack width. This is true for all the explored values of $\kappa/\mu$ except the smallest ones. This phenomenon is likely due to a change in the proportion between deviatoric and volumetric deformation, where a heavier penalty on volumetric strain energy is likely to reduce the driving force to crack propagation. Also, we should note that, in plane-strain conditions, the effective plane strain modulus is $E' = E/(1 - \nu^2)$, with $E$ the modulus in the material and $\nu$ the Poisson's ratio given by Eq. (4). Here, the correlation between $E'$ and $\nu$ is non-monotonic, and so is that between $E'$ and $\kappa/\mu$.

At low $G_c/\mu R$, we can observe that $a/R$ is almost independent of the strain-stiffening coefficient $\alpha$, and mainly controlled by the dimensionless bulk modulus $\kappa/\mu$, particularly for compressible materials. From this, we can deduce that volumetric deformation prevails over the deviatoric one, since given $a/R$ is highest in this regime and the needle is 'sinking' into the material (see the insets near the x-axis in Figure 3b). Here, $a/R$ is proportional to $\kappa/\mu$, due to an increased crack-opening energy for incompressible materials. Also, $a/R$ is proportional to $\alpha$ for the same reason.

Figure 3b shows that the dimensionless spacing (crack opening) energy $S_p/\mu R^2$ is proportional to the relative toughness $G_c/\mu R$ and, thus, inversely proportional to $a/R$, as shown in Eq. (9) and

Table 2. This is due to the larger energy required to open smaller cracks in tougher and/or softer materials.

At high $G_c/\mu R$, $S_p/\mu R^2$ appears to be mainly controlled by strain-stiffening $\alpha$, albeit this is mainly deduced from an extrapolation of the plots. This is mainly explained by the correlation between $S_p/\mu R^2$ and $a/R$, and suggests that deviatoric deformation prevails over the volumetric one. In this regime, $S_p/\mu R^2$ is proportional to both $\alpha$ and $\kappa/\mu$, since these parameters increase the energy required to open the crack.

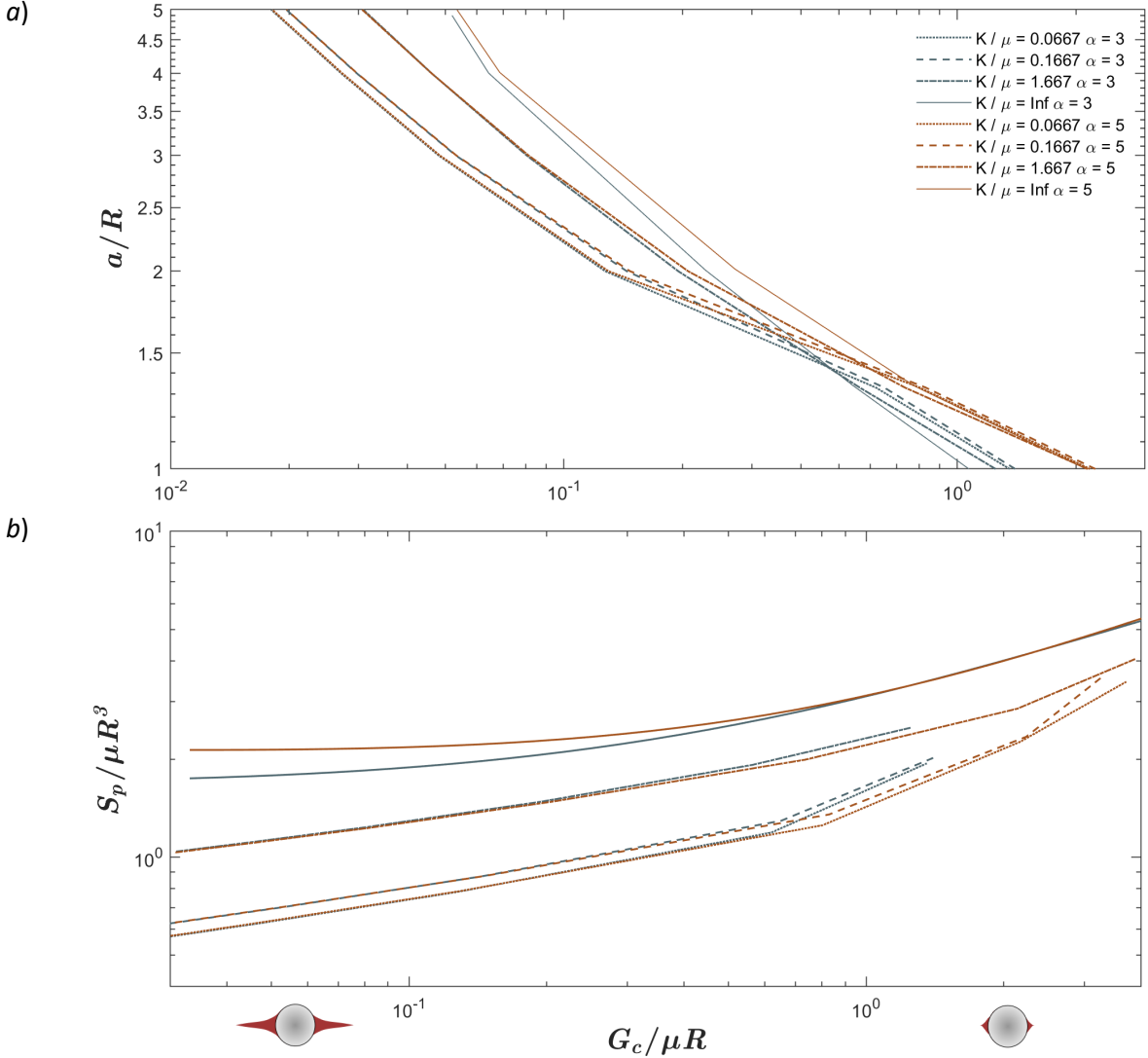

**Figure 3**: Dimensionless crack width $a/R$ ($R$ radius of the needle) a) and dimensionless spacing (crack opening) energy $S_p/\mu R^2$ ($\mu$ shear modulus) b) as a function of the *relative toughness* $G_c/\mu R$, the *strain-stiffening coefficient* $\alpha$, and the *dimensionless bulk modulus* $\kappa/\mu$ of the material

At low $G_c/\mu R$, $S_p/\mu R^2$ is mainly controlled by the dimensionless bulk modulus $\kappa/\mu$, as we observed for $a/R$. Also $S_p/\mu R^2$ is proportional to both $\kappa/\mu$ and $\alpha$. Here, the volumetric strain energy prevails again over the deviatoric one, due to the larger crack size, and we can make the same observations as in Figure 3a.

By substituting Eq. (8) and (9) into (7), with the fitting coefficients in Table 2, we can obtain the dimensionless penetration force $F_p/\mu R^2$ as a function of the relative needle radius $\mu R/G_c$ (the inverse of the relative toughness $G_c/\mu R$), the dimensionless bulk modulus $\kappa/\mu$, and the strain-stiffening coefficient $\alpha$ as plotted in Figure 4a.

Figure 4a shows that the dimensionless penetration force $F_p/\mu R^2$ is proportional to the relative toughness $G_c/\mu R$, and thus, inversely proportional to the relative needle radius (inverse of the relative toughness) $\mu R/G_c$.

At high $G_c/\mu R$, we can observe that $F_p/\mu R^2$ is inversely proportional to $\kappa/\mu$, suggesting that compressible materials are more puncture resistant than incompressible ones. This might be explained by the first term on the right-hand side of Eq. (7), which is proportional to $a/R$, while the latter is inversely proportional to $\kappa/\mu$, from Figure 3a. In this regime, we can also observe that $F_p/\mu R^2$ is inversely proportional to $\alpha$, for compressible materials, and directly proportional to $\alpha$, for incompressible materials. This is due to the second term on the right-hand side of Eq. (7), $S_p/\mu R^2$, which shows the same correlation with $\alpha$ in Figure 3b.

At low $G_c/\mu R$, we can observe that $F_p/\mu R^2$ is almost independent of strain stiffening $\alpha$, and, thus, is mainly controlled by the dimensionless bulk modulus $\kappa/\mu$. This was also observed for both the crack size $a/R$ and the spacing energy $S_p/\mu R^2$ in Figure 3, and these quantities contribute to the penetration force via Eq. (7). Here, $F_p/\mu R^2$ is proportional to both $\kappa/\mu$ and $\alpha$, as observed for $a/R$ and $S_p/\mu R^2$ in Figure 3, for all the explored values of $\kappa/\mu$, except the smallest one. In this case, incompressible materials exhibit higher puncture resistance than compressible ones, except for very small $\kappa/\mu$, where Figure 4a shows an added strength from compressibility. This result, however, is less reliable since low bulk moduli introduce a significant volumetric deformation, for which the linear approximation in Eq. (1c) and (2d) might result inadequate.

*Needle insertion*

As provided by [25], *indentation* is energetically favoured at shallower indentation depths, giving $w_i < w_p$ for $d < d_c$, while penetration becomes favoured at larger depths, thus $w_i > w_p$ for $d > d_c$. This finally provides that the critical energy required for needle insertion is $w_c = w_p^* = w_i^*$ for $d = d_c$, which, from Eq. (5) and (6), gives

$$\frac{d_c}{R} = \left(\frac{b+1}{B}\frac{F_p}{\mu R^2}\right)^{1/b} \tag{10}$$

Eq. (10) and the parameter values in Table 2 show that puncture-resistant materials, *i.e.* with high $F_p/\mu R^2$, also require deeper indentation to be pierced. By substituting Eq. (10) into (3) we can finally obtain the critical force $F_c$ required for needle insertion, giving

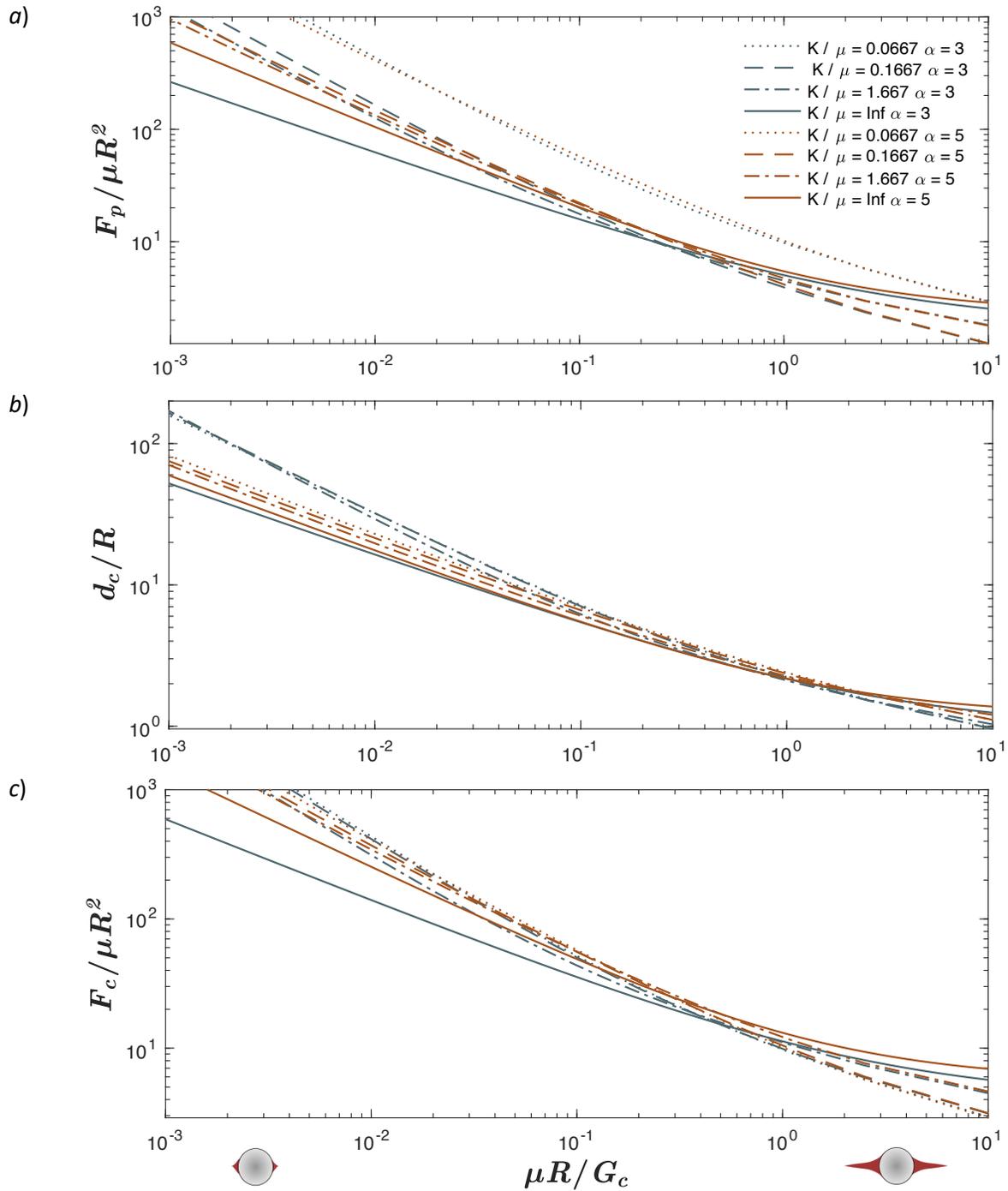

**Figure 4**: Dimensionless penetration force $F_p/\mu R^2$ *a*), critical indentation depth $d_c/R$ *b*), and force $F_c/\mu R^2$ *c*) at needle insertion, as a function of the *strain-stiffening coefficient* $\alpha$, the *dimensionless bulk modulus* $\kappa/\mu$, and the *relative toughness* $G_c/\mu R$ of the material (or the *relative radius* of the needle $\mu R/G_c$).

$$\frac{F_c}{\mu R^2} = (b+1)\frac{F_p}{\mu R^2} \tag{11}$$

Eq. (11) and Table 2 show here that puncture-resistant materials also require a larger force for needle insertion.

Taking Eq. (10) and (11), and substituting from Eq. (7), (8), and (9), with the parameter values from Tables 1 and 2, one can compute the dimensionless critical depth $d_c/R$ and force $F_c/\mu R^2$ - at needle insertion- as functions of the relative toughness $G_c/\mu R$, the strain-stiffening coefficient $\alpha$, and the dimensionless bulk modulus $\kappa/\mu$ of the material. These functions are plotted, respectively, in Figures 4b and 4c.

Figure 4b shows that the critical depth $d_c/R$ is proportional to the relative toughness $G_c/\mu R$, as so is the dimensionless penetration force $F_p/\mu R^2$ (Figure 4a), to which $d_c/R$ is proportional, via Eq. (10).

At high $G_c/\mu R$, $d_c/R$ is inversely proportional to $\kappa/\mu$, in agreement with our observations for Figure 4a. This implies that tough and/or soft materials require deeper indentation at needle insertion. However, this correlation is clear for high strain-stiffening, *i.e.* $\alpha = 5$, where for $\alpha = 3$ the above observation only applies for high bulk modulus. For $\alpha = 3$ and low bulk modulus we observe a direct proportionality between $d_c/R$ and $\kappa/\mu$, albeit in such a regime the linear-elastic approximation in Eq. (1c) and (2d) might introduce a significant error. In this regime, from Figure 4b, we can also observe that $d_c/R$ is proportional to $\alpha$ for incompressible materials, while $d_c/R$ is inversely proportional to $\alpha$ for compressible materials. The same was observed in Figure 4a for $F_p/\mu R^2$.

At low $G_c/\mu R$, $d_c/R$ is proportional to both $\kappa/\mu$ and $\alpha$, as previously observed for $F_p/\mu R^2$ in Figure 4a. This implies that brittle and/or stiff materials have the highest critical indentation depth at needle insertion when incompressible.

Figure 4c shows that also the critical force $F_c/\mu R^2$ is proportional to the relative toughness $G_c/\mu R$, as so is the dimensionless penetration force $F_p/\mu R^2$ (Figure 4a), to which $F_c/\mu R^2$ is proportional, via Eq. (11).

At high $G_c/\mu R$, $F_c/\mu R^2$ is inversely proportional to $\kappa/\mu$, in agreement with our observations for Figure 4a. This implies that tough and/or soft materials require a larger indentation force at needle insertion. In this regime, from Figure 4c, we can also observe that $F_c/\mu R^2$ is proportional to $\alpha$ for incompressible materials, while $F_c/\mu R^2$ is inversely proportional to $\alpha$ for compressible materials. The same was observed in Figure 4a for $F_p/\mu R^2$, and in Figure 4b, for $d_c/R$.

At low $G_c/\mu R$, $F_c/\mu R^2$ is proportional to both $\kappa/\mu$ and $\alpha$, as previously observed for $F_p/\mu R^2$ in Figure 4a. This implies that brittle and/or stiff materials have the highest critical indentation force at needle insertion when incompressible.

Eq. (2d) provides the linear elastic correlation between the hydrostatic pressure $p$ and the volumetric (engineering) strain $e_V = J - 1$ and, as discussed above, is likely to introduce an error. This error is proportionate to $e_V$, and thus to the dimensionless bulk modulus $\kappa/\mu$. Thus, our results are less accurate for the lowest values adopted for $\kappa/\mu$. Highly compressible materials, *i.e.* with low $\kappa/\mu$, will experience higher volumetric strain and, thus, the hydrostatic pressure $p$ should correlate nonlinearly with $J$, for better accuracy. One possibility would be to substitute Eq. (1c) with the series

$$\frac{\psi_V}{\mu} = \sum_{i=1}^{N} \frac{\kappa_i}{\mu} \frac{1}{2} (J-1)^{2i} \tag{12}$$

where $\kappa_i$ is the $i$-th bulk modulus. From Eq. (12), we should then substitute Eq. (2d) with

$$\frac{p}{\mu} = -\sum_{i=1}^{N} \frac{i\kappa_i}{\mu} (J-1)^{2i-1} \tag{13}$$

where now $p$ only correlates linearly with $J$ in for $i = 1$, while all other terms with $i > 1$ describe the nonlinear elastic response of the material to volumetric strain. However, this approach introduces more material parameters, thus rendering our study much more complicated.

## Conclusions

Our model predicts the puncture resistance of soft compressible solids from their mechanical properties. The puncture resistance is here defined as the penetration force $F_p$, and the indentation depth $d_c$ and force $F_c$ at needle insertion. Confirming previous findings, our model shows that tough and/or soft materials exhibit the highest puncture resistance. Surprisingly, our model shows that such materials exhibit the highest resistance when compressible, whereas incompressible materials provide minimal resistance. We speculate that the additional strength emerging from volumetric compressibility is due to the added volumetric strain energy stored in the material in the form of spacing energy, *i.e.*, the energy required to open the crack to accommodate the presence of the needle. We also observe that in brittle and/or stiff materials, exhibiting the lowest puncture resistance, volumetric compressibility detracts from puncture strength, leaving incompressible materials as the most resistant. Furthermore, we observe the interplay between volumetric compressibility and strain stiffening. The latter appears to increase puncture resistance in brittle/stiff materials while reducing resistance in tough/soft materials.

Finally, our theoretical findings suggest interesting strategies for the design of sharp needles and puncture-resistant materials. However, these findings require experimental validation, currently unavailable in the literature and, thus, left for future development.

## Acknowledgments
The work was supported by the Human Frontiers Science Program (RGY0073/2020), the Department of National Defense (DND) of Canada (CFPMN1–026), and the Natural Sciences and Engineering Research Council of Canada (NSERC) (RGPIN-2017–04464).